\documentclass[%
 reprint,
 amsmath,amssymb,
 aps,
prl,
floatfix,
]{revtex4-1}

\usepackage{ifpdf}
\usepackage{amsmath} 
\usepackage{amssymb}
\usepackage{amsfonts}
\usepackage{braket}
\usepackage{color}
\usepackage[colorlinks=true,linkcolor=blue,citecolor=blue,breaklinks]{hyperref}
\usepackage{graphicx}
\usepackage{dcolumn}
\usepackage{bm}
\usepackage[normalem]{ulem}

\newcommand{\la}{\langle}
\newcommand{\ra}{\rangle}

\usepackage{pdfpages} 
\makeatletter
\AtBeginDocument{\let\LS@rot\@undefined}
\makeatother

\begin{document}


\title{Local integrals of motion and the quasiperiodic many-body localization transition}
\author{Hansveer Singh$^1$, Brayden Ware$^1$, Romain Vasseur$^1$, and Sarang Gopalakrishnan$^{2,3}$}
\affiliation{$^1$Department of Physics, University of Massachusetts, Amherst, MA 01003, USA \\ $^2$Department of Physics, The Pennsylvania State University, University Park, PA 16820, USA \\ $^3$College of Staten Island, Staten Island, NY 10314, USA}


\begin{abstract}

We study the many body localization (MBL) transition for interacting fermions subject to quasiperiodic potentials by constructing the local integrals of motion (LIOMs) in the MBL phase as time-averaged local operators. We study numerically how these time-averaged operators evolve across the MBL transition. 
We find that the norm of such time-averaged operators drops discontinuously to zero across the transition; as we discuss, this implies that LIOMs abruptly become unstable at some critical localization length of order unity. 
We analyze the LIOMs using hydrodynamic projections and isolating the part of the operator that is associated with interactions. Equipped with this data we perform a finite-size scaling analysis of the quasiperiodic MBL transition. Our results suggest that the quasiperiodic MBL transition occurs at considerably stronger  quasiperiodic modulations, and has a larger correlation-length critical exponent, than previous studies had found.


\end{abstract}

\maketitle

Intuition suggests that isolated many-body systems initialized out of equilibrium should ``thermalize'' under their intrinsic unitary dynamics, in the sense of approaching a state in which \emph{local} observables and correlation functions exhibit equilibrium behavior~\cite{DeutschETH, SrednickiETH, Rigol07, Huse-rev}. Since Anderson's work~\cite{Anderson58}, it has been understood that thermalization is not fully generic: systems subject to strong quenched randomness can instead exhibit a many-body localized (MBL) phase, in which thermalization fails, and the system instead retains a local memory of its initial conditions to arbitrarily late times~\cite{Basko06, Mirlin05, BaskoMBLshort, OganesyanHuse, Huse-rev, AbaninRMP, imbrie2016many}. This memory is due to the existence in the MBL phase of (quasi-)local operators that are \emph{exact} integrals of motion, called LIOMs or l-bits~\cite{Serbyn13-1, Huse13, Kim14, Chandran14, ros2015integrals, imbrie2016many, imbrie2017local, gopalakrishnan2020dynamics}.
The existence of the MBL phase and of LIOMs has been established under minimal assumptions in one-dimensional random spin chains~\cite{imbrie2016many}; experimental evidence for the MBL phase exists in many different settings~\cite{Schreiber15, DeMarco15, smith2016many, Bloch16-2, bordia2017probing, Bordia17, roushan2017, xu2018, lukin2019probing, chiaro2019direct, rispoli2019quantum, leonard2020signatures, PhysRevB.99.035135, nguyen2020signature} (but see Refs.~\cite{Vidmar2019, vsuntajs2020ergodicity, sels2020dynamical, leblond2020universality, abanin2019distinguishing, panda2020can, crowley2020constructive}).
%
%
Rare regions---i.e., regions of a sample in which the disorder is anomalously weak or strong---play a central part in our understanding of MBL, determining the nature of the MBL transition~\cite{AltmanRG14, Potter15X, Dumitrescu17, Serbyn15, PhysRevLett.121.140601, GoremykinaPRL, MBLKT, MorningstarHuse, PhysRevB.102.125134, KhemaniCP}, response on both sides of the transition~\cite{Demler14, Reichman15, Gopa-15, Gopalakrishnan15Griffiths, Roeck17, luitz2017, PhysRevB.93.224201, PhysRevResearch.2.033262, agarwal2017rare, Luitz-subdiff, vznidarivc2016diffusive, Lezama19}, and even the stability of the MBL phase in higher dimensions~\cite{Roeck17, PhysRevB.99.134305}. However, many experiments on the MBL phase treat systems subject to \emph{quasiperiodic} (QP) rather than random potentials~\cite{Schreiber15, bordia2017probing, IyerQP, KhemaniCPQP, PhysRevB.96.104205, zhang2018universal, PhysRevB.98.224205, Doggen2019}. Noninteracting 1d QP systems have a localized phase~\cite{aubry1980analyticity}, which appears to be perturbatively stable in the presence of interactions~\cite{IyerQP}. However, rare regions are absent in QP systems, so it seems that the MBL transition---and the response near it---must differ qualitatively from the random transition~\cite{KhemaniCPQP}. The numerical evidence on the QP-MBL transition is mixed, with some studies casting doubt on whether a transition exists at all~\cite{vznidarivc2018interaction}, while others find a breakdown of diffusion~\cite{lev2017transport, PhysRevB.96.104205}, potentially even in a regime where single-particle states are delocalized~\cite{PhysRevLett.115.186601, PhysRevResearch.1.032039}.

Most work on QP-MBL systems has worked in the Schr\"odinger picture, considering the properties of typical individual eigenstates across the transition. 
The response of typical eigenstates in the MBL phase to a probe will involve both the external QP potential and self-generated \emph{configurational} randomness, from the random pattern of occupation of localized orbitals (which exert random Hartree shifts on one another). Thus from the eigenstate perspective there is no clear distinction between random and QP MBL systems; since the transition is really an instability of the MBL phase, one might be led to conclude that the transition should also be the same.  
In the present work, we instead take the \emph{Heisenberg} perspective and focus on the properties of the LIOMs as operators~\cite{agrawal2020universality}. 
In the QP-MBL phase, the pattern of LIOMs is quasiperiodic, with LIOMs approximately repeating at regular distances that are rational approximants to the QP pattern; there are no rare regions with anomalous LIOMs. 
Thus from this operator perspective the QP and random MBL phases differ, and one would also expect the transition at which LIOMs cease to exist to differ, if rare regions are indeed important. 
(Whether this transition coincides with the transition into the thermal phase is an issue we revisit below.) 

\begin{figure*}[!t]
	\includegraphics[width=0.95\textwidth]{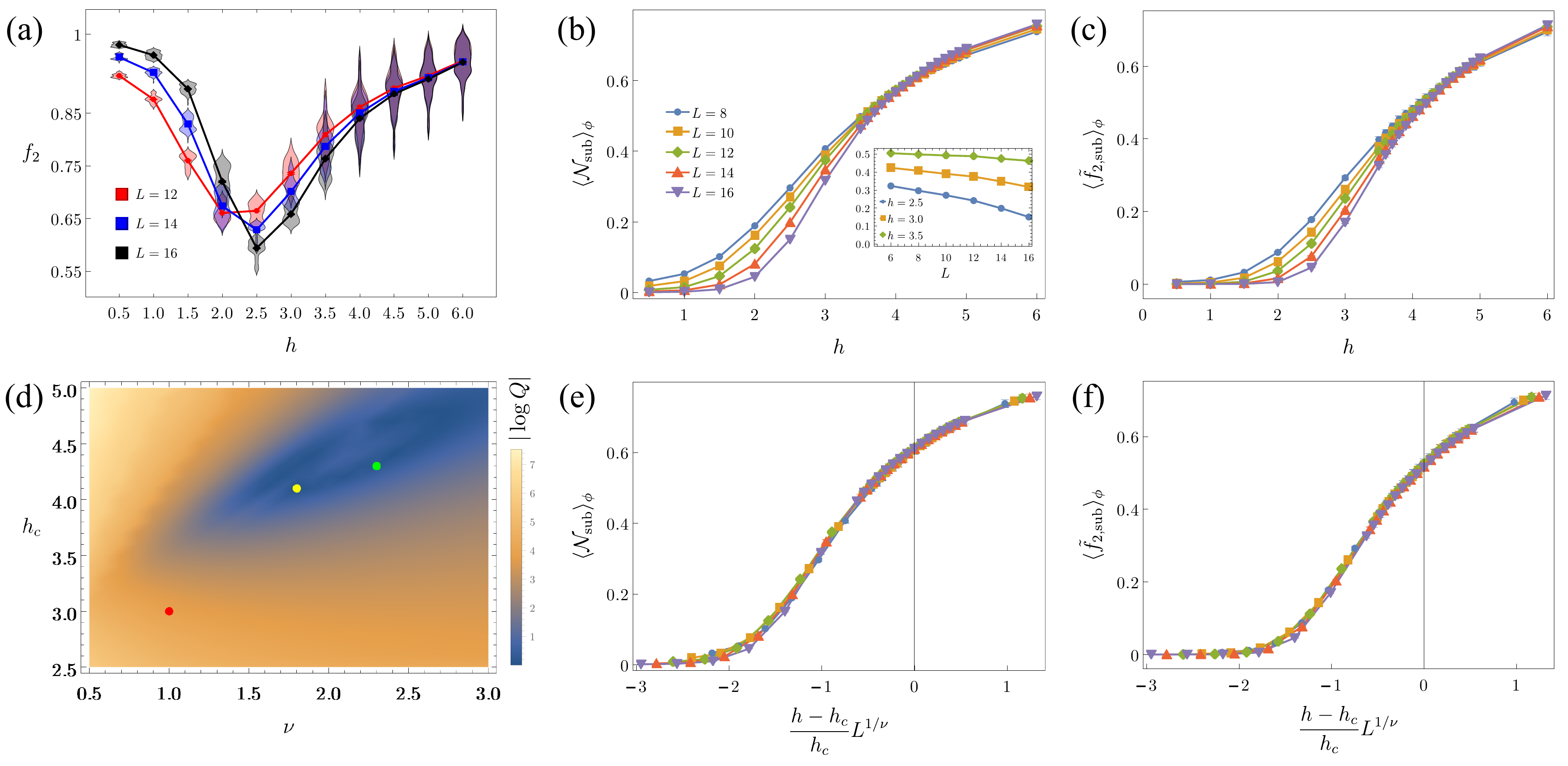}
	\caption{(a)~Evolution of the fraction $f_2$ of the weight of the time-averaged operator $\bar O$ that comes from fermion bilinears, vs. QP potential $h$. The solid lines are the average over the phase offset $\phi$ and the ``violin'' shapes indicate the distribution over $\phi$. (b)~Subtracted norm $\mathcal{N}_{\mathrm{sub}}$ of $\bar O$ vs. $h$, where the subtracted operator  is defined by projecting out hydrodynamic modes, eq.~\eqref{osub}. Note the crossing at $h \approx 4$. Inset shows the evolution of $\mathcal{N}_{\mathrm{sub}}$ with system size, indicating an instability that sets in as the system size is increased, even for $h = 3.5$ where previous studies have seen an MBL phase. (c)~Two-body component of the subtracted operator, defined in Eq.~\eqref{ftilde}; once again, this shows a crossing that is consistent with a discontinuous jump at the MBL transition. (d)~Quality of the finite-size scaling collapse of the curves in (b, c) as a function of the chosen collapse parameters $h_c, \nu$. The yellow dot marks the minimum of the cost function; by eye, the best collapse is for slightly different parameters (green dot) [these collapses are shown in panels (e) and (f)]. The transition seen in level statistics is marked by the red dot; our data are clearly inconsistent with these values. For all figures, averaging was performed over 200 phases equally spaced in the interval $[0,2\pi)$. The color scheme for different system sizes is shared for figures (b), (c), (e) and (f).}
	\label{fig:mainfig}
\end{figure*}

In the present work we explicitly construct LIOMs in QP systems by time-averaging local operators, as first proposed in Ref.~\cite{Chandran14}. We perform the infinite-time average explicitly, via full diagonalization (we also explore tensor-network methods~\cite{suppmat}). We analyze these LIOMs by computing the fraction of the operator norm that comes from $n$-fermion terms in the expansion $O = A_{ij} c^\dagger_i c_j + B_{ijkl} c^\dagger_i c^\dagger_j c_k c_l + \ldots$, using tensor-network methods to efficiently extract these quantities~\cite{suppmat}. These $n$-fermion weights give us a handle on the specifically \emph{many-body} effects that occur at the transition: unlike transport and entanglement, it is not contaminated by the single-particle critical point, which lies somewhat near the apparent many-body transition. We find that the $n$-fermion weights and the norm of the LIOMs give us new ways of analyzing the transition, pointing to a transition that occurs at larger values of the QP potential, with different critical exponents, than previously expected. This transition has notable similarities to the random case: in particular, the LIOMs slightly on the MBL side of the transition are tightly localized. 
Thus, as in the random case, it seems that the QP-MBL transition is an instability of the localized phase, which sets in at some critical value of the localization length.
The microscopic origin of this instability remains unclear.

{\it{Model.}--}
We consider the following model,
\begin{equation}\label{eqn:HXXZ}
H = \sum_{i=1}^{L-1}\sigma^x_{i}\sigma^x_{i+1}+\sigma^y_{i}\sigma^y_{i+1}+V \sigma^z_{i}\sigma^z_{i+1}+\sum_{i=1}^{L}h_{i}\sigma^z_{i},
\end{equation}
where $\sigma^{x,y,z}$ denote the Pauli matrices and $h_{i} = h \cos (2\pi/\varphi (i-L/2)+\phi)$ where $\varphi = \frac{1+\sqrt{5}}{2}$ and $\phi$ is a phase offset that we tune to translate our window. When $V=0$, this is the noninteracting Aubry-Andr\'e model which has localized eigenstates for $h>2$ and extended eigenstates for $h<2$. At nonzero $V$, finite size exact-diagonalization studies~\cite{IyerQP, KhemaniCPQP, PhysRevB.96.104205} of the average eigenstate entanglement and level statistics ratio have found an MBL transition at $h_{c}\approx 3$  with a critical exponent $\nu \sim 1$. (However, studies on longer spin-chains using the time-dependent variational principle have seen a larger critical point, consistent with ours~\cite{Doggen2019}.) In this letter we set $V = 1/2$.
\par
Following~\cite{Chandran14}, we construct LIOMs for this model by time averaging a local operator $O$, which we choose to be $\sigma^z_{L/2}$. The time average of $O$ is given by
\begin{equation}\label{timeaverage}
\bar{O} \equiv \lim_{T\to \infty} \frac{1}{T}\int_{0}^{T}dt ~O(t) =  \sum_{E} \la E | O | E \ra | E \ra\la E |,
\end{equation}
where $|E \ra$ are eigenstates of $H$. In the MBL phase, we expect $\bar O$ to be an approximately local operator with exponential tails, i.e., we expect there is some operator $O_n$ with support on $n$ sites such that $\Vert O - O_n \Vert \leq e^{-n/\xi}$ where $\xi$ is a characteristic localization length. In the ergodic phase, the time average instead produces a non-local integral of motion, predominately the projection of $O$ onto conserved charges.
We construct $\bar O$ by full exact diagonalization, using Eq.~\eqref{timeaverage}. We explore finite-time averages, performed using exact diagonalization as well as matrix-product methods, in~\cite{suppmat}.

{\it Fermion weights.}---
We analyze the LIOMs by expanding them in a basis of $n$-fermion operators. These are related to the Pauli operators by a Jordan-Wigner transformation, and evidently form a complete basis: 
\begin{equation}
\bar O = \sum_{\bm{\alpha}}c_{\bm{\alpha}}w_{1}^{\alpha_{1}}w_{2}^{\alpha_{2}}\cdots w_{2L}^{\alpha_{2L}},
\end{equation}
where $\{w_{i},w_{j}\} = 2\delta_{ij}$ are Majorana fermions. In what follows we will focus on two quantities: the Frobenius norm of the operator, $\mathcal{N} \equiv \mathrm{Tr} (\bar O^2)$, and its two-body weight $f_2$, defined by 
\begin{equation}
f_{2} = \frac{1}{\mathcal{N}}\sum_{|\bm{\alpha}|=2}|c_{\bm{\alpha}}|^{2},~\quad \mathcal{N} = \sum_{\bm{\alpha}}|c_{\alpha}|^{2}.
\end{equation}
We can also define four- and six-body weights accordingly. As $f_2$ is 1 for quadratic fermion operators, $1 - f_2 = f_4 + f_6 + \ldots$ measures the many-body content of the operator.
(These weights can be efficiently computed using matrix-product operator methods; we outline these methods and present results on the weights $f_n$ with $n>2$ in~\cite{suppmat}.) Note that $\mathcal{N}$ and $f_2$ probe complementary aspects of the time evolution: $\mathcal{N}$ addresses how much of the initial-state information survives in the time average, while $f_2$ addresses what fraction of this information is encoded in ``simple'' (i.e., few-body) operators.

{\it Hydrodynamic modes}.---
Fig.~\ref{fig:mainfig}(a) shows the average and distribution of $f_{2}$. 
As we might expect, $f_{2}$ approaches 1 and we also find that $\mathcal{N}$ is of order unity in the MBL phase, since in this phase the LIOMs are approximately single-site occupation numbers. 
Less obviously, $f_2$ also approaches 1 deep in the thermal phase (although $\mathcal{N}$, not shown, goes to zero with system size). One can understand this as follows. The operator $O = Z_i$ has some overlap with the total conserved charge, $I_1 \equiv \sum_i Z_i$, which is conserved (and is a two-fermion operator), and also with the Hamiltonian $I_2=H$ (which contains two- and four-point operators). 
More generally, in a system of size $L$, there are $2^L$ nonlocal conserved operators, i.e., projectors onto eigenstates, while the operator Hilbert space is $4^L$-dimensional. Since the operator $O$ at late times under chaotic dynamics is essentially random, its projection onto the conserved eigenstates would be exponentially small in $L$ if it were not for local conservation laws. 
Neglecting these exponentially small components, one can write $\bar O$ in the thermal phase as its projection onto hydrodynamic modes using the (super)projector 
\begin{equation}
\mathcal{P} = \sum_{l,k=1,2} |I_{k}\rangle\rangle C_{kl}^{-1} \langle\langle I_{l}|,
\end{equation}
where $I_1 = \sum_i Z_i$ and $I_2=H$ the conserved charges, acting on a Hilbert space $\mathcal{H}$, are now viewed as states on the doubled Hilbert space  $\mathcal{H} \otimes \mathcal{H}$, and $C_{kl} = \langle \langle I_{k} | I_{l} \rangle\rangle$ the susceptibility matrix with $\langle \langle A|B \rangle\rangle \equiv 2^{-L} \text{tr}(A^{\dagger}B)$. Since $H$ and $Q$ are both composed of two- and four-body operators, $f_2$ remains of order unity throughout the thermal phase. 

Since the hydrodynamic modes exist on both sides of the transition and the projection of an operator onto these modes is a property of the $t = 0$ operator that is insensitive to critical properties, we subtract the projection onto this hydrodynamic subspace and define the ``subtracted operator''
\begin{equation}\label{osub}
\bar{O}_{\text{sub}} \equiv \frac{\bar{O}-\mathcal{P}(O)}{\Vert O - \mathcal{P}(O) \Vert}.
\end{equation}
The denominator in Eq.~\eqref{osub} corrects for the fact that the hydrodynamic projection of $O$ smoothly increases with increasing disorder, since the Hamiltonian is dominated by single-site potential terms. (Empirically, we find that not fixing the normalization of $\bar{O}_{\mathrm{sub}}$ leads to spurious finite-size drifts in small-system numerics.) 
The norm $\mathcal{N}_{\mathrm{sub}}$ is defined as for the full operator. We also introduce the subtracted two body component, $\tilde{f}_{2 \mathrm{sub}}$, defined as 
\begin{equation}\label{ftilde}
\tilde{f}_{2,\mathrm{sub}} \equiv \frac{\sum_{i<j}|\text{tr}(\bar O_{\mathrm{sub}}w_{i}w_{j})|^{2}}{\sum_{i<j}|\text{tr}(O_{\mathrm{sub}}(0)w_{i}w_{j})|^{2}}.
\end{equation}
The rationale behind this normalization is, once again, to correct for the changes in the four-fermion weight of the subtracted operator as a function of $h$.

\emph{Results}.---Our results are summarized in Fig.~\ref{fig:mainfig}. Fig.~\ref{fig:mainfig}(a) shows the two-body weight $f_2$ of the full time-averaged operator (which, as noted above, is always $O \equiv \sigma^z_{L/2}$); as we expect, this is non-monotonic because it is dominated by hydrodynamic modes on the thermal side of the transition and by single-site operators deep on the MBL side of the transition. The sample-to-sample (or, equivalently, site-to-site) fluctuations of this quantity are large and size-independent deep in the MBL phase, negligible deep in the thermal phase, and intermediate in magnitude near the transition. Note that there is clear finite-size drift of $f_2$ for fields as large as $h = 4$, which previous literature~\cite{IyerQP, KhemaniCPQP, PhysRevB.96.104205} has assumed to be deep in the MBL phase. 

We now turn to the properties of the subtracted operator~\eqref{osub}. Fig.~\ref{fig:mainfig}(b) shows its norm, which decreases with system size in the thermal phase but saturates in the MBL phase. The decrease in the thermal phase is consistent with an exponential~\cite{suppmat}, which is what we would expect since we projected out hydrodynamic contributions. Similarly, the subtracted two-body component $\tilde f_{2, \mathrm{sub}}$ decreases continuously in the thermal phase and saturates in the MBL phase: this is, again, expected since the residual finite-size contributions to $\bar O_{\mathrm{sub}}$ in the thermal phase are highly nonlocal and have negligible two-body components.

While both quantities vanish identically in the thermodynamic limit throughout the thermal phase, it is not \emph{a priori} obvious whether they should rise continuously from zero or jump discontinuously at the MBL transition. Our numerical results strongly suggest the latter: the curves for both $\mathcal{N}_{\mathrm{sub}}$ and $f_{2, \mathrm{sub}}$ vs. system size cross in the interval $h \in (4, 4.5)$; moreover, the crossing shifts weakly to larger $h$ with increasing system size, suggesting that there are relatively ``simple'' LIOMs (with large two-body component) all the way up to the transition. This is consistent with a picture where the QP-MBL phase becomes unstable to thermalization at some critical value of the QP potential, but remains deeply localized all the way up to the transition. In the random case, ``avalanche''-based theories of the MBL transition generally predict this behavior, and it is also consistent with the available numerical evidence~\cite{luitz2017,ponte2017,PhysRevB.99.134205,PhysRevResearch.2.042033,PhysRevResearch.2.033262}. However, since avalanches do not obviously occur in QP systems, this behavior is unexpected (and had not previously been numerically observed to our knowledge). 

Another unexpected feature of these results is that the transition point and its critical properties differ quite strongly from those seen in previous numerical studies [Fig.~\ref{fig:mainfig}(d-f)]. Collapsing our finite-size data to the single-parameter scaling form $\phi(L, h - h_c) = \phi(L^{1/\nu} |h - h_c|/h_c)$, where $\nu$ is the correlation-length exponent, we find that the data collapses well in the parameter range $h_c \in (4, 4.5)$, with values of $\nu \in (2, 3)$. These results are very different from the expectation (gleaned, e.g., from studying the level statistics) that $h_c \approx 3$ and $\nu \approx 1$. In Fig.~\ref{fig:mainfig}(d) we have plotted the figure of merit [specifically, the log quality factor $|\log Q|$ extracted from the Python package pyfssa~\cite{andreas_sorge_2015_35293}] for attempted data collapses with various possible combinations of $h_c$ and $\nu$: our numerical data for the transition in the LIOMs are evidently inconsistent with previous predictions for the critical point. If anything, our scaling collapses show weak drift to larger values of $h_c$, suggesting that the transition might occur even deeper in the apparent MBL phase than our estimates above.

A clue as to why our results look so different from previous studies can be gleaned from the inset to Fig.~\ref{fig:mainfig}(b). For $h = 3, 3.5$ (which would conventionally be regarded as critical and localized respectively) $\mathcal{N}_{\mathrm{sub}}$ remains large for all the system sizes we study. However, it decreases with system size in a way that is \emph{accelerating} at larger $L$. This pattern is not consistent with the expected finite-size effects in the MBL phase, which should scale as $e^{-L/\xi}$, where $\xi$ is the correlation length, and should therefore flatten out at larger sizes. Rather, these results support a scenario in which the system seems localized on short scales, but then (beyond some critical scale $\xi$) realizes it is unstable. The scale $\xi$ diverges as $h$ is increased toward $h_c$, and can be regarded as a correlation length.



{\it Discussion}.---
In this work we studied the properties of LIOMs across the quasiperiodic many-body localization transition, constructing them as time-averaged local operators. In the thermal phase, these time-averaged operators are just projectors onto global conserved quantities like the total energy and charge; once these hydrodynamic parts are subtracted out, the remainder of the operator vanishes rapidly. In the MBL phase, instead, a time-averaged local operator retains a finite norm, since it has non-hydrodynamic projections onto the LIOMs. There is a transition at which these LIOMs cease to exist and the norm of the subtracted time-averaged operator vanishes. This apparent transition has a critical point and critical exponents that are inconsistent with the apparent transition in observables such as eigenstate entanglement and nearest-neighbor level statistics. Notably, the apparent correlation-length exponent $\nu \in (2, 2.5)$ that we extract from the finite-size scaling of the LIOMs is much larger than the Luck bound $\nu \geq 1$ (whereas previous results had $\nu = 1$, saturating the Luck bound). 
Indeed, we should emphasize that the results we have presented do not constitute strong evidence for the existence of an MBL phase at all, and are in principle consistent with a transition that occurs at $h_c = \infty$; however, the MBL phase is perturbatively stable for sufficiently large $h$, and no nonperturbative instabilities have yet been identified, so we take the point of view that there is a transition in the window where we see one. 
A counterintuitive implication of our results is that if $\nu \geq 2$, the QP-MBL critical point is \emph{perturbatively stable} against weak randomness by the Harris criterion~\cite{Harris}. 
(We note that a similar result was found using a real-space RG scheme in Ref.~\cite{zhang2018universal}.) 
This could suggest, either that there is a critical value of randomness required to change the universality class of the QP-MBL transition, or that both the QP-MBL critical point and some part of the QP-MBL phase undergo a nonperturbative instability for infinitesimal randomness.

How can we reconcile our observations with the results on level statistics and entanglement? One possibility is that there are two separate transitions with distinct critical properties, one at which the level statistics changes its character and another at which LIOMs cease to exist. This could happen, for example, if there were an intermediate phase with a many-body mobility edge~\cite{Alet14, Schiulaz15}. However, it is unclear whether such many-body mobility edges can exist~\cite{Schiulaz15}, and even if they do, the transition in entanglement should occur once the entire spectrum is localized. Thus it is not clear how this scenario could apply to our case. A second possibility is that the LIOMs we study here have weaker finite-size effects than the level statistics, because they are less affected by state-to-state fluctuations that are large in finite systems~\cite{KhemaniCPQP}. 
(All known finite-size effects favor the MBL phase, so a higher $h_c$ value is more plausible, assuming there is a single transition.) 

Our results shed light on the nature of this transition at which LIOMs cease to exist, which we tentatively identify with the MBL transition. In particular, we find that LIOMs even slightly on the MBL side are mostly fermion bilinears with large norm; thus they overlap strongly with microscopic spins. The QP transition, like the random one, appears to be an \emph{instability} of the MBL phase that sets in at some critical localization length as one increases the system size. In random systems, such an instability is thought to be seeded by rare regions that are locally thermal. Although rare regions do not exist, strictly speaking, in QP systems, one might still expect the instability to occur first in some parts of the sample. One might expect LIOMs to be unusually delocalized in samples that contain these parts. However, we do not see much evidence of enhanced heterogeneity at the transition (Fig.~\ref{fig:mainfig}). It therefore seems that the instability we are seeing is due to the proliferation of many-body resonances in \emph{typical} regions of the sample. The origin of these resonances remains to be identified. 

Our work, like all ED studies, is inherently limited to small system sizes. An important question for future work is whether one can construct LIOMs for much larger systems. We attempted to do so by time evolving local operators via time-evolving block decimation (TEBD) applied to matrix-product operators~\cite{suppmat}, and averaging over finite time windows. Unfortunately, to get a good approximation to the LIOMs away from the deeply localized limit, one must average over such long time windows (comparable to the Heisenberg time) that TEBD is impractical~\cite{suppmat}, because the bond dimension needed to describe the operator grows intractably large. Whether other forms of explicit time-averaging, e.g., based on Krylov-space methods~\cite{Luitz-subdiff, PhysRevB.100.104303}, can provide access to larger systems and sufficiently long times is an interesting question for future work. 

\begin{acknowledgments}
The authors thank U. Agrawal, P. Dumitrescu, D. Huse, V. Khemani and V. Oganesyan for helpful discussions. S.G. acknowledges support from NSF DMR-1653271. R.V.  acknowledges support from the US Department of Energy, Office of Science, Basic Energy Sciences, under Early Career Award No. DE-SC0019168 and the Alfred P. Sloan Foundation through a Sloan Research Fellowship.

\end{acknowledgments}

\bibliography{mbl}

\bigskip

\onecolumngrid
\newpage



\section*{Supplementary material}

# Supplemental Material: Local integrals of motion and the quasiperiodic many-body localization transition

Hansveer Singh[1], Brayden Ware[1], Romain Vasseur[1], and Sarang Gopalakrishnan[2,3]

[1]*Department of Physics, University of Massachusetts, Amherst, MA 01003, USA*
[2]*Department of Physics, The Pennsylvania State University, University Park, PA 16820, USA*
[3]*College of Staten Island, Staten Island, NY 10314, USA*
(Dated: January 10, 2021)

## I. COMPUTATION OF THE $k$-BODY FERMION WEIGHT

Here we provide a formula to compute the $k$-body fermion weight. To start we write an operator, $O$, in the basis of Majorana fermions

$$O = \sum_{\boldsymbol{\alpha}} c_{\boldsymbol{\alpha}} w_1^{\alpha_1} w_2^{\alpha_2} \cdots w_{2L}^{\alpha_{2L}}, \tag{S1}$$

where $w_i$ are hermitian operators which satisfty $\{w_i, w_j\} = 2\delta_{ij}$. The $k$-body fermion weight is given by,

$$f_k = \frac{1}{\mathcal{N}} \sum_{|\boldsymbol{\alpha}|=k} |c_{\boldsymbol{\alpha}}|^2, \quad \mathcal{N} = \sum_{\boldsymbol{\alpha}} |c_{\boldsymbol{\alpha}}|^2. \tag{S2}$$

It is straightforward to show that,

$$\sum_{|\boldsymbol{\alpha}|=k} |c_{\boldsymbol{\alpha}}|^2 = \frac{1}{4^L} \sum_{|\boldsymbol{\alpha}|=k} |\text{tr}(O^\dagger w_1^{\alpha_1} w_2^{\alpha_2} \cdots w_{2L}^{\alpha_{2L}})|^2, \tag{S3}$$

where $|\boldsymbol{\alpha}| \equiv \sum_i \alpha_i$, and it is also straightforward to see that

$$\sum_{\boldsymbol{\alpha}} |c_{\boldsymbol{\alpha}}|^2 = \frac{1}{2^L} \text{tr}(O^\dagger O). \tag{S4}$$

Plugging in the above relations into Eq. (S2), one finds

$$f_k = \frac{1}{2^L \text{tr}(O^\dagger O)} \sum_{|\boldsymbol{\alpha}|=k} |\text{tr}(O^\dagger w_1^{\alpha_1} w_2^{\alpha_2} \cdots w_{2L}^{\alpha_{2L}})|^2. \tag{S5}$$

## II. CONSTRUCTING L-BITS WITH TEBD

As discussed in the main text the l-bits are obtained through time averaging, i.e.

$$\bar{O} = \lim_{T \to \infty} \frac{1}{T} \int_0^T dt\, O(t). \tag{S6}$$

$\bar{O}$ is expected to be approximately local (starting from a local operator) deep in the MBL phase, so $\bar{O}$ can be efficiently represented as an MPO [S1]. Our main text suggests that deep in the thermal phase $\bar{O}$ is dominated by the hydrodynamic modes of our system and so it should be well described by an MPO since the only (dominant) hydrodynamic modes are $H$ and $M$. Near the transition it is less clear how accurate an MPO description is of $\bar{O}$. Denoting $\xi$ as the characteristic length over which the l-bits have support (this is only well defined in the MBL phase) then intuition from the random case suggests that $\xi$ should remain finite as one approaches the delocalization transition from the MBL phase (see Section VII). Motivated by this idea we attempted to construct the l-bits using TEBD.

### A. Method Description

Proceeding with the discussion of $\bar{O}$'s construction, we define the finite time average $\bar{O}(T) \equiv \frac{1}{T}\int_0^T dt O(t)$. Discretizing time into intervals of $\Delta t$ and denoting $M \equiv T/\Delta t$, then

$$\bar{O}(T) \approx \frac{1}{M} \sum_{n=1}^{M} O(n\Delta t) + \mathcal{O}((\Delta t)^2). \tag{S7}$$

Using this approximation we have the following relation,

$$\bar{O}(T) = \frac{1}{M}((M-1)\bar{O}(T - \Delta t) + O(T)). \tag{S8}$$

Keeping track of the finite time average, we can use this formula to construct $\bar{O}(T)$ by adding the two MPO representations of $\bar{O}(T - \Delta t)$ and $O(T)$. If $\bar{O}(T - \Delta t)$ is described by a bond dimension $\psi$ MPO and $O(T)$ is described by a bond dimension $\chi$ MPO then $\bar{O}(T)$ is described by a bond dimension $\chi + \psi$ MPO. Since the enlarged bond dimension, $\chi + \psi$, may not be the optimal bond dimension for $\bar{O}(T)$ or it may be too large to handle numerically, one generally needs a compression scheme to lower the bond dimension.

The standard method of compressing an MPO is done by applying an SVD to each tensor in the MPO and discarding the appropriate number of singular values until either we are below some error threshold, $\epsilon$, or above some maximal bond dimension, $D$. This method will give the optimal tensor of bond dimension $D$ to use for a given site but this may not be the best global representation of $\bar{O}$ at bond dimension $D$. An improvement to this method is to use an iterative compression scheme [S2] which takes an initial guess for the globally best MPO and then minimizes the Frobenious norm between the truncated average, $\bar{O}(T)_{\text{approx}}$, and the exact average, i.e. we iteratively minimize $||\bar{O}(T) - \bar{O}(T)_{\text{approx}}||^2$ where $||A||^2 = \text{tr}(A^\dagger A)$.

Let us summarize the above steps: we calculate $O(T)$, then using $\bar{O}(T - \Delta t)$ we construct $\bar{O}(T)$ using Eq. (S8) and using the SVD compressed approximation of $\bar{O}(T)$ as our initial guess, we employ iterative compression to bring $\bar{O}(T)$ down to a lower bond dimension.

To perform the standard TEBD step we pick $\Delta t = 0.05$ and use a fourth order trotterization scheme of the time evolution operator given by [S3, S4],

$$e^{-iH\Delta t} \approx U(\tau_1)U(\tau_1)U(\tau_2)U(\tau_1)U(\tau_1) + \mathcal{O}((\Delta t)^5), \tag{S9}$$

where

$$U(t) = e^{-iH_{\text{odd}}t/2}e^{-iH_{\text{even}}t}e^{-iH_{\text{odd}}t/2}, \tag{S10}$$

$\tau_1 = \frac{\Delta t}{4 - 4^{1/3}}$, $\tau_2 = \Delta t - 4\tau_1$ and $H_{\text{even/odd}}$ are the terms in the Hamiltonian acting on even/odd bonds in the chain. As we perform time evolution we keep track of the operator norm square of $\bar{O}(T)$ and the two body weight of $\bar{O}(T)$. While the former quantity is easily computable via standard matrix-product techniques, the fact that $f_2$ can be efficiently computed is less trivial. Not only do we show how this can be done in Section IV for $f_2$ but also for all $f_k$ where $k$ is even.

### B. Results

Unfortunately we found two big computation walls to this method: convergence time of observables and bond dimension. In Fig. (S1) we demonstrate that already at $L = 8$ we see a long convergence time is required for $f_2$ and when we are closer to the transition this time scale increases. Furthermore, we see that when we are deeper into the MBL phase we have fairly good convergence with bond dimension-seeing convergence occuring close to $\chi = 32$. On the other hand when we are near the transition even for $\chi = 64$ (25% of the Hilbert space dimension) we are unable to converge at long times which suggests that we cannot get near the transition if we were to increase system size. In addition we remark that the inaccuracy that develops in the description of $\bar{O}(T)$ is due to the truncation error from $O(T)$ which is a well known issue in performing long time simulations with TEBD. We also note that it may be the case that $\bar{O}$ still has an approximate MPO representation but it is not practical to construct it using TEBD. Thus we find that using matrix-product methods is not a practical tool to study the MBL transition but could be used to examine quantities deep within the MBL phase as previous studies have done [S1, S5]. We will however use MPO techniques to compute the $k$-body weight even for system sizes accessible to exact diagonalization (see below).

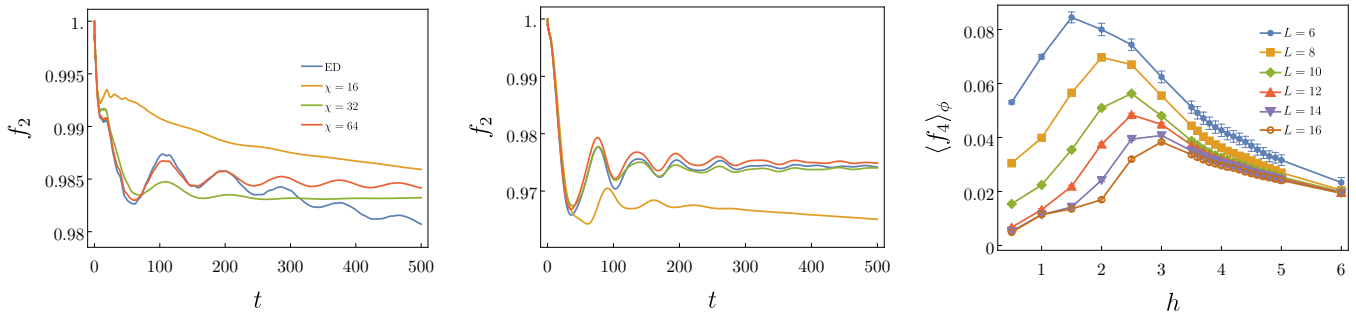

FIG. S1. The (*left*) and (*middle*) plots show the time evolution for $f_2$ with $L = 8$, zero global phase shift, and $h = 4.5$ or $h = 5$ respectively. It is evident from these two plots that convergence with bond dimension is exceedingly better when we are deeper inside the MBL phase rather than near the transition. The color scheme for both the (*left*) and (*middle*) are the same. (*right*) The four body weight averaged over 200 phases uniformly spaced in the interval $[0, 2\pi)$ and the error bars signify the standard deviation. Clear strong finite size effects can be seen in $f_4$ and based on the maximum of $f_4$ the critical point seems to be closer to $h \approx 3$.

## III.  ED COMPUTATION OF $f_k$

While MPOs are not a viable route to study the full phase diagram, the MPO construction of $f_k$ turned out to be a useful way to compute $f_k$ using ED. The standard brute force method would require using Eq. (S5) which is not very efficient even though the $w_i$ are sparse matrices. An improvement to this would be to utilize the $U(1)$ symmetry present in the problem and compute the overlap with normal ordered fermion operators $c_i^\dagger c_j$ , $c_i^\dagger c_j^\dagger c_k c_l$., etc. Another way we found that was not extremely expensive was to utilize Eq. (S16) (in the next section) and create $O_+$ by arranging all the blocks in a $2^L \times 2^L$ matrix and build $H_k$ by contracting the tensors. The cost of matrix multiplication is $\mathcal{O}(4^L(2k+1)^2)$ but in practice we found that diagonalization was still the dominant computation.

In Fig. (S1) we present the average of $f_4$ and its behavior as a function of quasiperiodic modulation. Contrary to $\tilde{f}_{2,\mathrm{sub}}$, this suffers from much larger finite size effects, which accounts for the smaller critical point value. Further finite size effects do not weaken if we consider its subtracted counterpart (not shown), $f_{4,\mathrm{sub}}$, which is defined in Section V.

## IV.  EXTENDING THE $k$-BODY WEIGHT CALCULATION FOR MATRIX PRODUCT OPERATORS

Here we show that the $k$-body weight can be efficiently computed for matrix product operators. We first write an operator, $O$, in the pauli basis,

$$O = \sum_{\boldsymbol{\beta}} b_{\boldsymbol{\beta}} \sigma_1^{\beta_1} \sigma_2^{\beta_2} \cdots \sigma_L^{\beta_L}. \tag{S11}$$

Using the Jordan-Wigner transformation,

$$w_{2j-1} = \prod_{k<j} Z_k X_j \ , \quad w_{2j} = -\prod_{k<j} Z_k Y_j, \tag{S12}$$

one can show that Majorana strings and pauli strings are related to each other by a phase, i.e. $w_1^{\alpha_1} w_2^{\alpha_2} \cdots w_{2L}^{\alpha_{2L}} = e^{i\theta(\boldsymbol{\alpha})} S(\boldsymbol{\alpha})$. Applying this fact to the numerator of Eq. (S5), we see that

$$\sum_{|\boldsymbol{\alpha}|=k} |\mathrm{tr}(O^\dagger w_1^{\alpha_1} w_2^{\alpha_2} \cdots w_{2L}^{\alpha_{2L}})|^2 = \sum_{|\boldsymbol{\alpha}|=k} |\sum_{\boldsymbol{\beta}} b_{\boldsymbol{\beta}}^* \mathrm{tr}(\sigma_1^{\beta_1} \sigma_2^{\beta_2} \cdots \sigma_L^{\beta_L} S(\boldsymbol{\alpha}))|^2$$

$$= 4^L \sum_{|\boldsymbol{\alpha}|=k} |b_{\boldsymbol{\beta}(\boldsymbol{\alpha})}|^2.$$

Going to from the first to the last line, I used the orthonormality of pauli strings and used $\boldsymbol{\beta}(\boldsymbol{\alpha})$ to denote the pauli string corresponding to the Majorana string with occupation number $\boldsymbol{\alpha}$. The above calculation illustrates another way we can compute the numerator of Eq. (S5). Defining the two operators,

$$O_+ = \sum_{\boldsymbol{\beta}} |b_{\boldsymbol{\beta}}|^2 \sigma_1^{\beta_1} \sigma_2^{\beta_2} \cdots \sigma_L^{\beta_L}, \tag{S13}$$

and

$$H_k = \sum_{|\boldsymbol{\alpha}|=k} S(\boldsymbol{\alpha}), \tag{S14}$$

one can see that

$$\sum_{|\boldsymbol{\alpha}|=k} \text{tr}(O_+^\dagger S(\boldsymbol{\alpha})) = 2^L \sum_{|\boldsymbol{\alpha}|=k} |b_{\boldsymbol{\beta}(\boldsymbol{\alpha})}|^2, \tag{S15}$$

which shows

$$f_k = \frac{\text{tr}(O_+^\dagger H_k)}{\text{tr}(O^\dagger O)}. \tag{S16}$$

Next, we show that Eq. (S16) can be computed efficiently when $O$ and $H_k$ have MPO representations. Writing down the MPO representation of $O$ in the pauli basis,

$$O = \sum_{\boldsymbol{\beta},\boldsymbol{a}} A^{\beta_1[1]}_{a_0,a_1} A^{\beta_2[2]}_{a_1,a_2} \cdots A^{\beta_L[L]}_{a_{L-1},a_L} \sigma_1^{\beta_1} \sigma_2^{\beta_2} \cdots \sigma_L^{\beta_L}, \tag{S17}$$

one can see that

$$|b_{\boldsymbol{\beta}}|^2 = \sum_{\boldsymbol{a},\boldsymbol{a}'} (A^{\beta_1[1]}_{a_0,a_1} A^{\beta_1[1]*}_{a_0',a_1'}) \cdots (A^{\beta_L[L]}_{a_{L-1},a_L} A^{\beta_L[L]*}_{a_{L-1}',a_L'}). \tag{S18}$$

Denoting the tensors for the MPO representation of $H_k$ as $W^{\beta_j[j]}_{a_{j-1}'',a_j''}$ and using the above expression one finds

$$\text{tr}(O_+^\dagger H_k) = \sum_{\boldsymbol{\beta},\boldsymbol{\beta}',\boldsymbol{a},\boldsymbol{a}',\boldsymbol{a}''} (A^{\beta_1[1]}_{a_0,a_1} W^{\beta_1'[1]}_{a_0'',a_1''} A^{\beta_1[1]*}_{a_0',a_1'}) \cdots (A^{\beta_L[L]}_{a_{L-1},a_L} W^{\beta_L'[L]}_{a_{L-1}'',a_L''} A^{\beta_{L-1}[L]*}_{a_{L-1}',a_L'}) \text{tr}(\sigma_1^{\beta_1}\sigma_1^{\beta_1'}\sigma_2^{\beta_2}\sigma_2^{\beta_2'}\cdots\sigma_L^{\beta_L}\sigma_L^{\beta_L'})$$

$$= 2^L \sum_{\boldsymbol{\beta},\boldsymbol{a},\boldsymbol{a}',\boldsymbol{a}''} (A^{\beta_1[1]}_{a_0,a_1} W^{\beta_1[1]}_{a_0'',a_1''} A^{\beta_1[1]*}_{a_0',a_1'}) \cdots (A^{\beta_L[L]}_{a_{L-1},a_L} W^{\beta_L[L]}_{a_{L-1}'',a_L''} A^{\beta_{L-1}[L]*}_{a_{L-1}',a_L'})$$

$$= 2^L \sum_{\boldsymbol{\beta},\boldsymbol{\beta}',\boldsymbol{a},\boldsymbol{a}',\boldsymbol{a}''} (A^{\beta_1[1]}_{a_0,a_1} W^{\beta_1'[1]}_{a_0'',a_1''} \delta^{\beta_1,\beta_1'} A^{\beta_1'[1]*}_{a_0',a_1'}) \cdots (A^{\beta_L[L]}_{a_{L-1},a_L} W^{\beta_L'[L]}_{a_{L-1}'',a_L''} \delta^{\beta_L,\beta_L'} A^{\beta_{L-1}'[L]*}_{a_{L-1}',a_L'}).$$

The last line illustrates that this is an expectation value of a superoperator MPO whose tensors are given by $\mathcal{W}^{\beta_j\beta_j'[j]}_{a_{j-1},a_j} = W^{\beta_j'}_{a_{j-1}'',a_j} \delta^{\beta_j,\beta_j'}$ and so the complexity of the calculation scales polynomially with the bond dimension of $O$ and $H_k$. In the following subsection we show that $H_k$ can be represented by an MPO whose bond dimension only depends on $k$.

### A. MPO construction of $H_k$

In order to construct the MPO representation of $H_2$, we need to first identify all the Majorana strings with $|\boldsymbol{\alpha}|=2$. Using Eq. (S12), the two body fermion operators up to a phase are given by,

$$X_j \prod_{i=j+1}^{k-1} Z_i Y_k, \quad Y_j \prod_{i=j+1}^{k-1} Z_i Y_k, \quad X_j \prod_{i=j+1}^{k-1} Z_i X_k, \quad Y_j \prod_{i=j+1}^{k-1} Z_i X_k, \quad Z_j, \tag{S19}$$

where $X, Y, Z$ denote the standard Pauli matrices. We can now use the method described in Ref. [S6], where one constructs a weighted finite automaton to generate the operator whose summands are those above. The diagram which constructs this weighted finite automaton is given in Fig. (S2).

To construct the MPO for $H_4$ we have to identify all the Majorana strings with $|\boldsymbol{\alpha}|=4$. To find such strings, it is useful to introduce the following diagrammatic notation for the Majorana strings:

$$| \longleftrightarrow \text{single site } X_j \text{ or } Y_j$$

$$\text{———} \longleftrightarrow \text{strings of } Z_j \text{ with length greater than or equal to 2}, \prod_j Z_j$$

$$\times \longleftrightarrow \text{single site } Z_j$$

$$\circ \longleftrightarrow \text{single site identity}, \mathbb{1}_j.$$

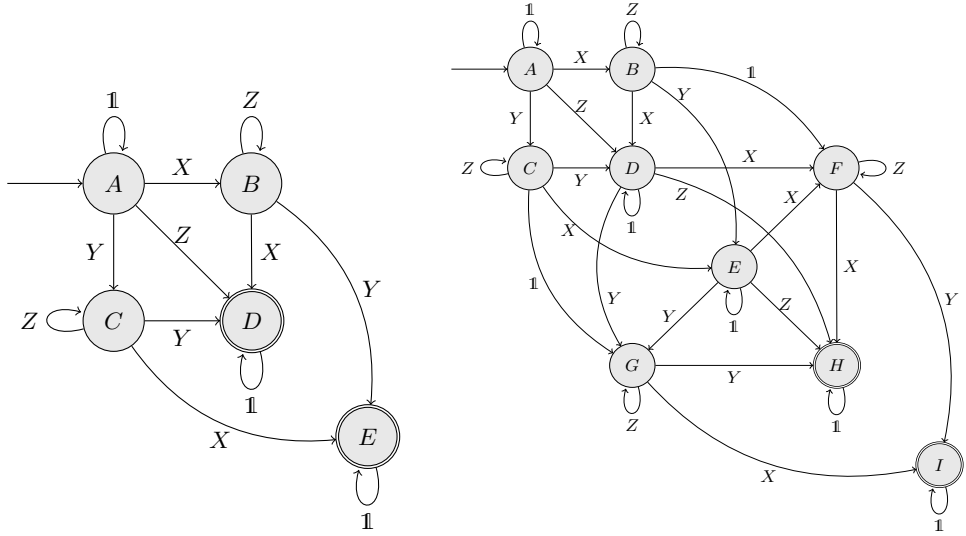

FIG. S2. Diagrams which construct the weighted finite automata for the (*left*) two body hamiltonian and (*right*) the four body hamiltonian.

Using this diagrammatic notation and Eq. (S19), one finds that only two diagrams correspond to the two body operators:

$$\vdash\!\!\!\dashv \longleftrightarrow X_j \prod_{i=j+1}^{k-1} Z_i Y_k, \ Y_j \prod_{i=j+1}^{k-1} Z_i Y_k, \ X_j \prod_{i=j+1}^{k-1} Z_i X_k, \ Y_j \prod_{i=j+1}^{k-1} Z_i X_k$$

$$\times \longleftrightarrow Z_j.$$

This suggests the following interpretation: two vertical lines connected by a horizontal line contributes two fermion operators and an "X" contributes two fermion operators. One can now write down the diagrams for the four body operators,

$$\vdash\!\!\!\dashv \ \vdash\!\!\!\dashv \longleftrightarrow \text{two non-overlapping two body strings}$$
$$\times \ \vdash\!\!\!\dashv \longleftrightarrow \text{a single site } Z_j \text{ with a non-overlapping two body string}$$
$$\times \ \times \longleftrightarrow \text{two non-overlapping single site } Z_j$$
$$\vdash\!\!\!\text{O}\!\!\!\dashv \longleftrightarrow \text{a two body string broken by an insertion of a } Z_j.$$

The last diagram tells us that a circle contributes two fermion operators while a vertical line with a horizontal line attached contributes one fermion operator. Accounting for the possible spatial orderings of the diagrams and the fact that a vertical line can either be $X_j$ or $Y_j$, one arrives at the diagram that constructs the weighted finite automaton shown in Fig. (S2).

The diagrams can be used to construct the matrices as shown in Ref. [S6]. Denoting $W_{2/4}^{\sigma_j \sigma'_j [j]}$ the matrices in the

MPO representation of $H_{2/4}$, then

$$W_2^{\sigma_1 \sigma_1'[1]} = \begin{pmatrix} \mathbb{1} & X & Y & Z & 0 \end{pmatrix}, \tag{S20}$$

$$W_2^{\sigma_j \sigma_j'[j]} = \begin{pmatrix} \mathbb{1} & X & Y & Z & 0 \\ 0 & Z & 0 & X & Y \\ 0 & 0 & Z & Y & X \\ 0 & 0 & 0 & \mathbb{1} & 0 \\ 0 & 0 & 0 & 0 & \mathbb{1} \end{pmatrix}, \quad 1 < j < L, \tag{S21}$$

$$W_2^{\sigma_L \sigma_L'[L]} = \begin{pmatrix} Z \\ X+Y \\ X+Y \\ \mathbb{1} \\ \mathbb{1} \end{pmatrix}, \tag{S22}$$

and

$$W_4^{\sigma_1 \sigma_1'[1]} = \begin{pmatrix} \mathbb{1} & X & Y & Z & 0 & 0 & 0 & 0 & 0 \end{pmatrix}, \tag{S23}$$

$$W_4^{\sigma_j \sigma_j'[j]} = \begin{pmatrix} \mathbb{1} & X & Y & Z & 0 & 0 & 0 & 0 & 0 \\ 0 & Z & 0 & X & Y & \mathbb{1} & 0 & 0 & 0 \\ 0 & 0 & Z & Y & X & 0 & \mathbb{1} & 0 & 0 \\ 0 & 0 & 0 & \mathbb{1} & 0 & X & Y & Z & 0 \\ 0 & 0 & 0 & 0 & \mathbb{1} & X & Y & Z & 0 \\ 0 & 0 & 0 & 0 & 0 & Z & 0 & X & Y \\ 0 & 0 & 0 & 0 & 0 & 0 & Z & Y & X \\ 0 & 0 & 0 & 0 & 0 & 0 & 0 & \mathbb{1} & 0 \\ 0 & 0 & 0 & 0 & 0 & 0 & 0 & 0 & \mathbb{1} \end{pmatrix}, \quad 1 < j < L, \tag{S24}$$

$$W_4^{\sigma_L \sigma_L'[L]} = \begin{pmatrix} 0 \\ 0 \\ 0 \\ Z \\ Z \\ X+Y \\ X+Y \\ \mathbb{1} \\ \mathbb{1} \end{pmatrix}. \tag{S25}$$

One can see in Fig. (S2) that the four body diagram is essentially two two body diagrams (the first two body diagram is with states $A-D$ and the second is formed with states $E-I$) joined together by extra transitions between states. This extends to arbitrary $k$ body weight for $k$ even and the matrices for the $k$ body weight with $k$ even have bond dimension $2k+1$ and the initial and final MPO tensors are the same but with 0s padding to the right of the first tensor and the top of the last tensor. The bulk MPO tensor is given by repeating the first four rows of eq. (S24) for $2k-4$ times and then the last $5 \times 5$ block is given by the the corner $5 \times 5$ block in the lower right corner of eq. (S24).

## V. COMPUTATION OF THE $k$-BODY WEIGHT FOR THE SUBTRACTED L-BIT

Here we calculate the $k$-body weight of the subtracted l-bit. Given a Hamiltonian with conserved quantities, $I_k$, which are not necessarily linearly independent, then the (super)projector onto the subspace spanned by the $I_k$ is given by

$$\mathcal{P} = \sum_{lk} |I_k\rangle\rangle C_{kl}^{-1} \langle\langle I_l|, \tag{S26}$$

where the $I_k$, acting on a Hilbert space $\mathcal{H}$, are now viewed as states on the Hilbert space $\mathcal{H} \otimes \mathcal{H}$ and $C_{kl} = \langle\langle I_k | I_l \rangle\rangle$ with $\langle\langle A|B\rangle\rangle \equiv 2^{-L} \text{tr}(A^\dagger B)$. Using Eq. (S5), the $k$-body weight for the subtracted l-bit, denoted as $f_{k,\text{sub}}$ is given by,

$$f_{k,\text{sub}} = \frac{1}{2^L \text{tr}(\bar{Z}_{\text{sub}}^2)} \sum_{|\boldsymbol{\alpha}|=k} |\text{tr}(\bar{Z}_{\text{sub}} w_1^{\alpha_1} w_2^{\alpha_2} \cdots w_{2L}^{\alpha_{2L}})|^2, \quad \bar{Z}_{\text{sub}} = \bar{Z}_k - \mathcal{P}(\bar{Z}_k). \tag{S27}$$

In this section we heavily abuse notation and will usually replace $\bar{Z}_k$ with $\bar{Z}$ and $Z_k$ with $Z$. We first calculate the subtracted norm, i.e. the denominator of Eq. (S27).

$$\begin{aligned}
\text{tr}(\bar{Z}_{\text{sub}}^2) &= \text{tr}(\bar{Z}^2) + \text{tr}(\mathcal{P}(\bar{Z})^2) - 2\,\text{tr}(\bar{Z}\mathcal{P}(\bar{Z})) \\
&= \text{tr}(\bar{Z}^2) + \text{tr}(\mathcal{P}(\bar{Z})^2) - 2\,\text{tr}((\mathcal{P}(\bar{Z}) + \mathcal{Q}(\bar{Z}))\mathcal{P}(\bar{Z})) \\
&= \text{tr}(\bar{Z}^2) - y^T C^{-1} y.
\end{aligned}$$

Going from the second to the third line we wrote $\bar{Z} = \mathcal{P}(\bar{Z}) + \mathcal{Q}(\bar{Z})$ where $\mathcal{Q}$ is the (super)projector onto the orthogonal complement of the hydrodynamic subspace and the fact that $\langle\langle I_k|\bar{Z}\rangle\rangle = \langle\langle I_k|Z\rangle\rangle$. In the last line we introduced the vector $y_k \equiv \langle\langle I_k|Z\rangle\rangle$. Using the fact that $H$ and the magnetization, $M \equiv \sum_i Z_i$, are the only (dominant) hydrodynamic modes, we find

$$C_{HM} = \sum_{k=1}^{L} h_k, \quad C_{HH} = 2(L-1) + V^2(L-1) + \sum_{k=1}^{L} h_k^2, \quad C_{MM} = L, \tag{S28}$$

$$y_H = h_k, \quad y_M = 1. \tag{S29}$$

Before calculating the numerator we remark that since $\mathcal{P}(\bar{Z})$ only contains operators with only two or four body fermion operators (in our model) then the numerator of with $\bar{Z}_{\text{sub}}$ is the same as $\bar{Z}$. Thus we only need to calculate the numerator for $k=2$ and $k=4$.

### A. Computation of the subtracted 2-body weight

Straightforward computation shows that,

$$\begin{aligned}
\sum_{|\boldsymbol{\alpha}|=2} |\text{tr}(\bar{Z}_{\text{sub}} w_1^{\alpha_1} \cdots w_{2L}^{\alpha_{2L}})|^2 &= \sum_{i<j} |\text{tr}(\bar{Z}_{\text{sub}} w_i w_j)|^2 \\
&= \sum_{i<j} |\text{tr}(\bar{Z} w_i w_j)|^2 + |\text{tr}(\mathcal{P}(\bar{Z}) w_i w_j)|^2 - 2\,\text{Re}(\text{tr}(\bar{Z} w_i w_j)^* \text{tr}(\mathcal{P}(\bar{Z}) w_i w_j)).
\end{aligned} \tag{S30}$$

The second term can also be straightforwardly computed,

$$\begin{aligned}
\sum_{i<j} |\text{tr}(\mathcal{P}(\bar{Z}) w_i w_j)|^2 &= \sum_{i<j} |\sum_{kl} C_{kl}^{-1} \langle\langle I_l|Z\rangle\rangle \text{tr}(I_k w_i w_j)|^2 \\
&= \sum_{klmn} C_{kl}^{-1} C_{mn}^{-1} \langle\langle I_l|Z\rangle\rangle \langle\langle I_n|Z\rangle\rangle \sum_{i<j} \text{tr}(I_k w_i w_j)^* \text{tr}(I_m w_i w_j) \\
&= y^T C^{-1} F C^{-1} y,
\end{aligned} \tag{S31}$$

where $F_{km} = \sum_{i<j} \text{tr}(I_k w_i w_j)^* \text{tr}(I_m w_i w_j)$. To calculate $F_{km}$ we first write $H$ and $M$ in the Majorana basis. Using Eq. (S12), we get

$$H = \sum_{k=1}^{L-1} -iw_{2k} w_{2k+1} + iw_{2k-1} w_{2k+2} - Vw_{2k-1} w_{2k} w_{2k+1} w_{2k+2} - i\sum_{k=1}^{L} h_k w_{2k-1} w_{2k} \tag{S32}$$

$$M = \sum_{k=1}^{L} -iw_{2k-1} w_{2k}, \tag{S33}$$

and so

$$\text{tr}(H w_i w_j) = 2^L \left( \sum_{k=1}^{L-1} i\delta_{2k\,2k+1,ij} - i\delta_{2k-1\,2k+2,ij} + i\sum_{k=1}^{L} h_k \delta_{2k-1\,2k,ij} \right) \tag{S34}$$

$$\text{tr}(M w_i w_j) = 2^L \sum_{k=1}^{L} i\delta_{2k-1\,2k,ij}. \tag{S35}$$

Using the above results we can compute $F_{km}$ and find

$$F_{HM} = 4^L \sum_{k=1}^{L} h_k, \quad F_{HH} = 4^L(2(L-1) + \sum_{k=1}^{L} h_k^2), \quad F_{MM} = 4^L L. \tag{S36}$$

We now calculate the third term in Eq. (S30). Plugging in the expression for $\mathcal{P}(\bar{Z})$, we find

$$\sum_{i<j} \mathrm{Re}(\mathrm{tr}(\bar{Z}w_i w_j)^* \mathrm{tr}(\mathcal{P}(\bar{Z})w_i w_j)) = \sum_{lm} C_{lm}^{-1} y_l \mathrm{Re}(\sum_{i<j} \mathrm{tr}(\bar{Z}w_i w_j)^* \mathrm{tr}(I_m w_i w_j))$$

$$= -\sum_{lm} C_{lm}^{-1} y_l \mathrm{Re}(\sum_{i<j} \mathrm{tr}(\bar{Z}w_i w_j) \mathrm{tr}(I_m w_i w_j))$$

$$= -y^T C^{-1} z,$$

where $z_m = \mathrm{Re}(\sum_{i<j} \mathrm{tr}(\bar{Z}w_i w_j) \mathrm{tr}(I_m w_i w_j))$. Using Eq. (S34), we find that

$$z_H = 2^L \mathrm{tr}(\bar{Z}(-H_{\mathrm{free}}))$$

$$= 2^L(-\mathrm{tr}(ZH) + V \sum_{i=1}^{L-1} \mathrm{tr}(\bar{Z}Z_i Z_{i+1})), \tag{S37}$$

$$z_M = 2^L \mathrm{tr}(-ZM),$$

where $H_{\mathrm{free}}$ refers to the part of $H$ which only contains two body operators. Defining the vector $d_k = (V\mathrm{tr}(\bar{Z} \sum_{i=1}^{L-1} Z_i Z_{i+1}), 0)$, then $z_k = 2^L(d_k - y_k)$ and the third term in Eq. (S30) becomes,

$$\sum_{i<j} \mathrm{Re}(\mathrm{tr}(\bar{Z}w_i w_j)^* \mathrm{tr}(\mathcal{P}(\bar{Z})w_i w_j)) = 2^L y^T C^{-1} y - 2^L y^T C^{-1} d \tag{S38}$$

Plugging Eq. (S31) and Eq. (S38) into Eq. (S30), we get

$$\sum_{|\boldsymbol{\alpha}|=2} |\mathrm{tr}(\bar{Z}_{\mathrm{sub}} w_1^{\alpha_1} \cdots w_{2L}^{\alpha_{2L}})|^2 = \sum_{i<j} |\mathrm{tr}(\bar{Z}w_i w_j)|^2 + y^T C^{-1} F C^{-1} y - 2^{L+1}(y^T C^{-1} y - y^T C^{-1} d). \tag{S39}$$

### B. Computation of the subtracted 4-body weight

The calculation is carried out in a similar fashion as the 2-body weight.

$$\sum_{|\boldsymbol{\alpha}|=4} |\mathrm{tr}(\bar{Z}_{\mathrm{sub}} w_1^{\alpha_1} \cdots w_{2L}^{\alpha_{2L}})|^2 = \sum_{i<j} |\mathrm{tr}(\bar{Z}_{\mathrm{sub}} w_i w_j)|^2$$

$$= \sum_{i<j<k<l} |\mathrm{tr}(\bar{Z}w_i w_j w_k w_l)|^2 + |\mathrm{tr}(\mathcal{P}(\bar{Z})w_i w_j w_k w_l)|^2 \tag{S40}$$

$$- 2\,\mathrm{Re}(\mathrm{tr}(\bar{Z}w_i w_j w_k w_l)^* \mathrm{tr}(\mathcal{P}(\bar{Z})w_i w_j w_k w_l)).$$

The second term will have the same form as Eq. (S31), but with with a new $F$ which we call $F^4 = \sum_{i<j<k<l} \mathrm{tr}(I_k w_i w_j w_k w_l)^* \mathrm{tr}(I_m w_i w_j w_k w_l)$. Notice $F^4$ will only have $F_{HH}^4 \neq 0$. We find that

$$\mathrm{tr}(H w_i w_j w_k w_l) = -2^L \sum_{p=1}^{L-1} V \delta_{2p-1\,2p\,2p+1\,2p+1, ijkl}, \tag{S41}$$

thus

$$F_{HH}^4 = 4^L V^2 (L-1). \tag{S42}$$

The third term in Eq. (S40) also has the same form except there is no minus coming from the conjugate transpose of the four fermion operator and now the vector $z$ is redefined as $z^4$ and only $z_H^4 \neq 0$. We find,

$$z_H^4 = 2^L V \mathrm{tr}(\bar{Z} \sum_{i=1}^{L-1} Z_i Z_{i+1}), \tag{S43}$$

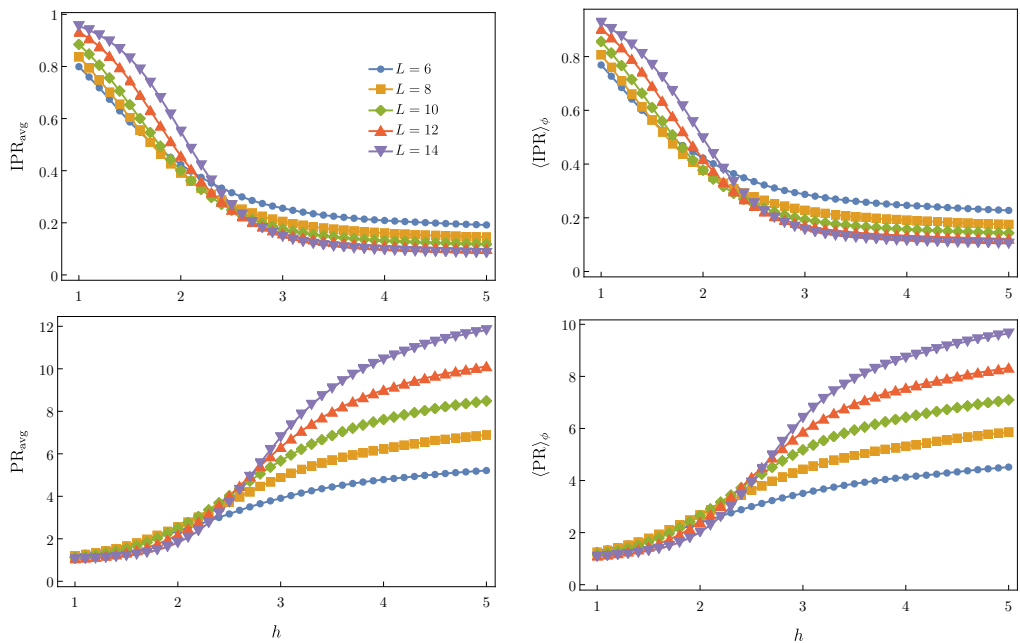

FIG. S3. (*top left*) IPR and (*bottom left*) PR of the singular values of $\langle \ell_{jk} \rangle_\phi$ averaged over 176 phase realizations uniformly distributed between $[0, 2\pi)$. The (*top right*) average IPR and (*bottom right*) average PR of the singular values $\ell_{jk}$.

which means $z_k^4 = d_k$. Putting the results together we have

$$\sum_{|\boldsymbol{\alpha}|=4} |\text{tr}(\bar{Z}_{\text{sub}} w_1^{\alpha_1} \cdots w_{2L}^{\alpha_{2L}})|^2 = \sum_{i<j<k<l} |\text{tr}(\bar{Z} w_i w_j w_k w_l)|^2 + y^T C^{-1} F^4 C^{-1} y - 2^{L+1} y^T C^{-1} d). \tag{S44}$$

## VI. LINEAR INDEPENDENCE OF THE L-BITS

The l-bits we used were constructed by performing time evolution on a local operator followed by a time average. These l-bits are different than the usual $\tau_i^{\alpha_i}$ which are orthonormal and only exist in the MBL phase. First, they are not necessarily orthonormal but as was shown in Ref. [S5] the overlaps between l-bits exponentially decay as one goes deep into the MBL phase for the random case. Second, our l-bits are conserved operators regardless of where we sit in the phase diagram. Here we numerically demonstrate that the l-bits are indeed linearly independent in the quasiperiodic case by computing the inverse participation ratio (IPR) and participation ratio (PR) of the singular values of the matrix $\ell_{jk} \equiv \langle\langle \bar{Z}_j | \bar{Z}_k \rangle\rangle$, i.e. $\text{IPR} = \sum_i s_i^4$ and $\text{PR} = 1/\text{IPR}$. In Fig. S3 we show the average IPR and average PR as well as the IPR and PR from the singular values of the averaged matrix $\langle \ell_{jk} \rangle_\phi$. Recall the number of non-zero singular values directly translates to the rank of $\ell_{jk}$ (or its averaged counterpart). This means deep in the thermal phase we expect the time averaged operators to be linearly dependent so the rank should be close to 1 and so the IPR should be close to 1. In the MBL phase we expect the time averaged operators to be linearly independent and so all the singular values should be of the same order. Since we normalize the singular values to satisfy $\sum_{i=1}^L s_i^2 = 1$ then each $s_i^2$ should approximately scale as $1/L$ and so the IPR should vanish in the MBL phase. As one can see in Fig. S3 both of these limiting behaviors are exhibited in the IPR and in the PR (since it is just the reciprocal of the IPR). In addition, a crossing seems to occur for $h \in (2, 3)$ which is considerably lower than the estimate in the main text but this is due to stronger finite size effects in the IPR than to our observables in the main text.

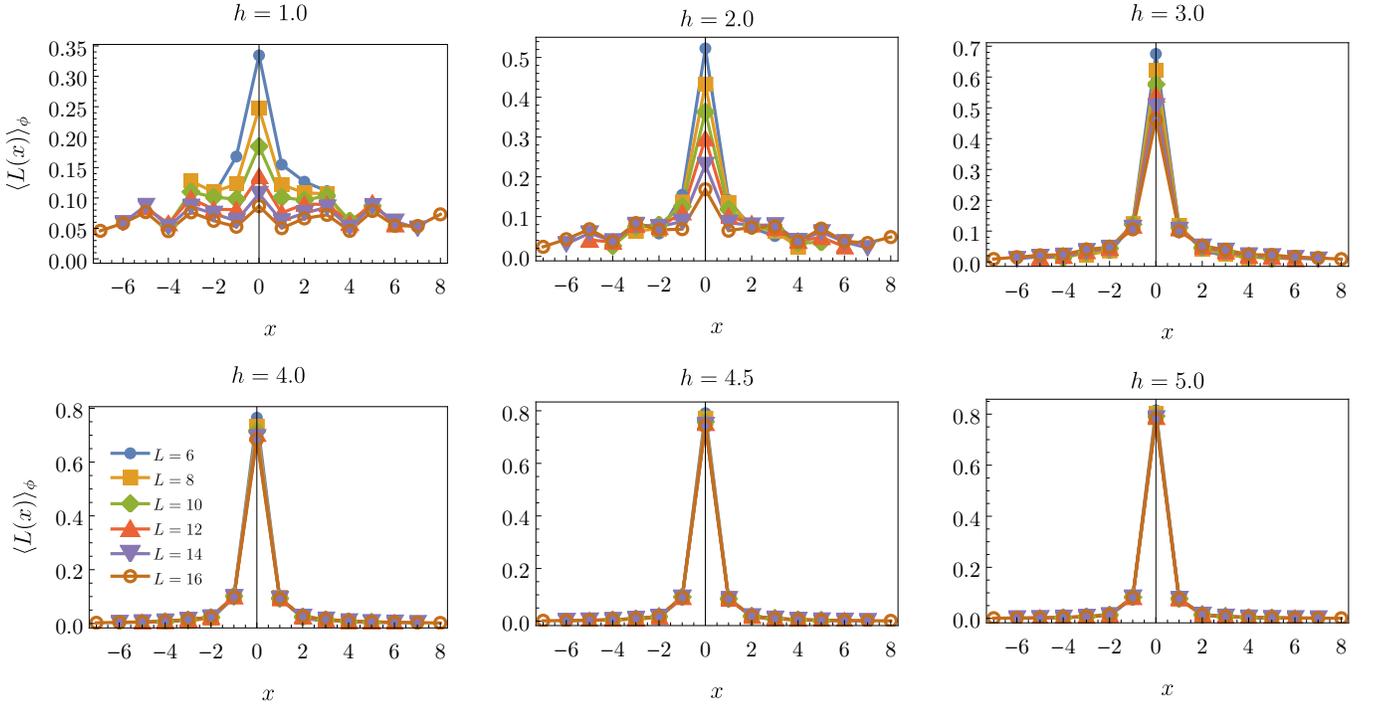

FIG. S4. Operator locator density averaged over 176 phase realizations uniformly distributed between $[0, 2\pi)$. $x$ is the lattice site shifted by half the system size so that all system sizes have the same origin. One can see that the l-bit remains fairly localized even as we are approaching the vicnity of the critical point given by either the subtracted norm or subtracted two body component.

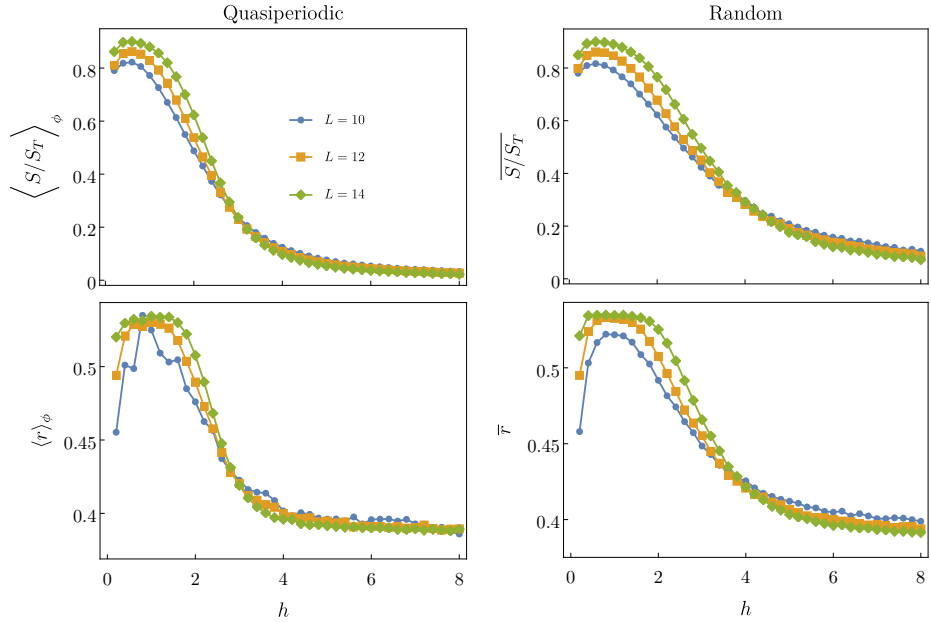

FIG. S5. [(*left*) and (*right*) Average half entanglement entropy divided by the page value, $S_T$, and the average level statistics ratio, $r$ for the quasiperiodic case (*left*) and the random case (*right*). One can see that a crossing occurs around $h \approx 3$ for the quasiperiodic case while the crossing occurs closer to $h \approx 4$ in the random case. The quasiperiodic case was averaged over 500 phase realizations and the random case was averaged over 1000 realizations. In both cases the average half entanglement entropy and the level statistics ratio were computed in the $M = 0$ sector removing 10% of the eigenstates from the extremes of the spectrum.

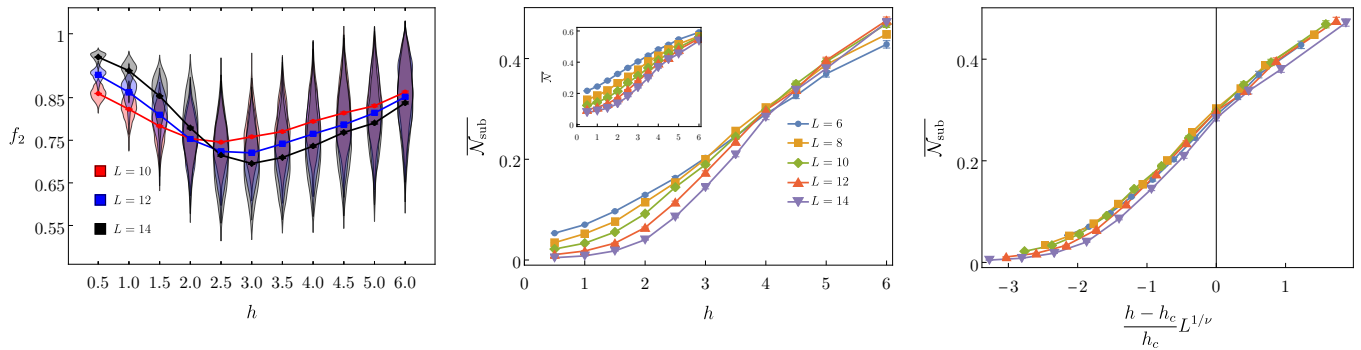

FIG. S6. (*left*) Distribution of $f_2$ of $\bar{Z}_{L/2}$ over 899 disorder realizations picked uniformly from $[-h, h]$. Solid lines indicate the average of $f_2$. One can see stronger finite size effects and broader distributions compared to the quasiperiodic case consistent with previous studies. (*middle*) Here is the average subtracted norm squared of $\bar{Z}_{L/2}$ which seems to exhibit a crossing near $h \approx 4.5$, but the number of realizations averaged over is too low to make any definitive prediction for $h_c$. The inset shows the unsubtracted norm squared of $Z_{L/2}$ which does not exhibit a crossing. Overall there is qualitatively similar behavior between the random case and the quasiperiodic case but quantitatively the subtracted norm squared is lower than the quasiperiodic case. (*right*) Here we perform a collapse of the subtracted norm squared data for $\nu = 2$. Although the low sample size does not make this estimate precise, we did find that $\nu = 1$ does not gives a worse collapse than $\nu = 2$.

## VII. SPATIAL STRUCTURE OF THE L-BITS

Here we present a characterization of the spatial structure of our l-bits. To do so, we decompose our l-bit in the Pauli basis given by Eq. (S11). One can then compute the left-right weight distribution given by,

$$\rho(z, y) = \frac{1}{\mathcal{N}} \sum_{|\boldsymbol{\beta}| = |z-y|} |b_{\boldsymbol{\beta}}|^2, \quad \mathcal{N} = \sum_{\boldsymbol{\beta}} |b_{\boldsymbol{\beta}}|^2, \tag{S45}$$

which gives us the fraction of weight the l-bit has in Pauli strings which start at site $z$ and end at site $y$. Finally, one can define another quantity called the operator locator density defined as,

$$L(x) = \sum_{z \leq x \leq y} \frac{\rho(z, y)}{y - z + 1}, \tag{S46}$$

which gives a weighted average of Pauli strings which have support on site $x$. This quantity gives a picture of where the support of the l-bit is. As can be seen in Fig. (S4), at large $h$ the l-bit only has support near the center of the chain since we constructed our l-bit from $Z_{L/2}$, while at small $h$ the l-bit has support across the whole chain since the l-bit becomes a random operator in the ergodic phase. As we go towards the transition $4 < h < 4.6$ from the MBL phase, the l-bit still remains localized near the center of the chain suggesting that the size of the l-bit remains finite even at the transition similar to the random case.

## VIII. COMPARISON TO THE RANDOM CASE

In this section we show how the random case deviates from the quasiperiodic case. We first show in Fig. (S5) the average half entanglement entropy, $S$, divided by the Page value for a random pure state, $S_T = 1/2(L \log 2 - 1)$, and the average level statistics ratio, $r \equiv \min(\Delta_n, \Delta_{n+1})/\max(\Delta_n, \Delta_{n+1})$ where $\Delta_n = E_n - E_{n+1}$ is the spacing between energy eigenvalues for the $M = 0$ sector. Examining Fig. (S5), we observe a crossing at $h \approx 3$ for the quasiperiodic case and $h \approx 4$ for the random case.

In Fig. (S6) we present our new observables $f_2$ and $\mathcal{N}_{\text{sub}}$. Similar to Refs. [S7, S8] we see broader distributions in $f_2$ compared to the quasiperiodic case as well as stronger finite size effects. While we only averaged over a small sample size of 899 disorder realizations, one can see that a crossing seems to occur at higher disorder strengths than previously seen before and with a critical exponent of $\nu \approx 2$ which satisfies the Chayes bound. Another feature is that the norm squared of the operator is smaller in the localized phase than in the quasiperiodic case which another

reassuring signal that the MBL transition sits at higher disorder strength.



\end{document}